# Focused Coherent Microwave Scattering for Spatially Resolved Electron Number Density Diagnostics

H. Jenrow, K. Reindersma, A. Shashurin

*School of Aeronautics and Astronautics, Purdue University, West Lafayette, IN, USA*

**This work provides an initial demonstration of Focused Coherent Microwave Scattering (F-CMS) diagnostics technique, an extension of the conventional Coherent Microwave Scattering (CMS) technique that enables spatially resolved measurements. The spatial-resolution functionality was achieved using PTFE lenses to focus the probing microwave beam and isolate a distinct probing volume. The feasibility of F-CMS was demonstrated using glow-discharge plasma volume with an electron number density of approximately $5\times10^9$ $cm^{-3}$ as a test object. A clear F-CMS output signal was successfully detected upon the discharge initiation. These findings establish a foundation for the further development of the F-CMS technique, which has the potential to enable highly sensitive, spatiotemporally resolved measurements of electron number density in real time, without the need for signal accumulation (sensitivity down to an electron number density of approximately $10^8$ $cm^{-3}$ is expected). Such capability will be useful for a variety of applications including diagnostics of unsteady flowfields such as those occurring in hypersonic shock tunnels and spacecraft electric propulsion systems.**

## 1. Introduction

Coherent microwave scattering (CMS) has been traditionally used for diagnostics of small plasma objects with sizes much smaller than the probing wavelength, in which case the plasma behaves as a point dipole.[1,2] The technique measures the total electron count in the plasma volume ($N_e$), and the electron number density ($n_e$) can be determined if plasma volume ($V_p$) is independently measured ($n_e = \frac{N_e}{V_p}$). The CMS technique has been successfully applied in a wide variety of applications including laser-induced plasmas,[3,4,5] nanosecond pulsed discharges,[6] non-equilibrium atmospheric pressure plasma jets,[7] microdischarges used for electrosurgery,[8,9] small-size glow discharges in glass enclosures,[10] and inductively coupled plasmas.[11]

The CMS technique is associated with outstanding sensitivity for $n_e$-measurements in real-time (down to $n_e$~$10^{12}$ $cm^{-3}$).[2] Fundamentally, this high sensitivity results from the fact that the scattering contributions from all elements within the small plasma volume under test add up in-phase at the detector, which is due to use of a large centimeter-scale microwave wavelength. The entire plasma volume effectively "sees" the same phase of the incident microwave field, and the detector is nearly equidistant from each plasma element. As a result, all wavelets produced by individual plasma elements add up in-phase at the detector and scattering is deemed constructive (in-phase) coherent.

Some recent efforts have sought to enable a spatial resolution in the CMS technique via using conductive tube enclosures to expose only a selected portion of the plasma to the probing microwaves.[5] Using this approach, one-dimensional mapping of infrared laser filaments at a wavelength of 3.9 µm was successfully demonstrated. However, this approach is intrusive due to required utilization of metallic enclosure tubes, and it cannot be applied to large-scale plasmas that cannot be enclosed within such tubes.

In this work, we demonstrate feasibility of achieving spatially resolved measurements of electron number density in large plasma volumes via utilization of PTFE lenses to focus the probing microwaves and isolate a distinct probing volume. A glow-discharge plasma with a known electron number density was used as a test plasma object.

## 2. Methodology

The proposed novel F-CMS technique was demonstrated via measurements of the electron number density in a glow-discharge facility, as shown in Figure 1(a). The chamber consisted of an acrylic tube with length of 35.6 cm and diameter of 20.4 cm. The chamber was evacuated using an Alcatel M2012A vacuum pump, and the air pressure was maintained at 0.3 torr (40 Pa) during all experiments. The two electrodes were thin conductive disks of 18.8 cm diameter, separated by 9.2 cm, with an area covering almost all of the cross section of the discharge tube. The discharge was sustained using the Spellman SL-2000 DC power supply at 630 V and 150 mA; a 6 kΩ resistor was placed in series with the glow discharge tube to stabilize the current, and the total voltage output of the DC power supply was 1530 V, with 900 V dropped across the resistor.

The F-CMS system was constructed around the glow discharge facility as shown in Figure 1(b). 14.525 GHz microwaves were emitted by a Herley PDRO-14794 source, and the signal was then passed through a Millitech AMC-15-RFH80 frequency multiplier, increasing the frequency to 58.1 GHz. An isolator was used to stop any reflected signals from traveling back to the emitter components. The microwave power after the isolator was measured at 25 dBm. The signal was then split into two arms using an M/A COM 507514 directional coupler: the sample and reference arms. The sample arm was taken from the directional-coupler output (~25 dBm), whereas the reference arm was taken from the 20-dB attenuated forward-wave coupling port (~5 dBm). The sample-arm signal was emitted with vertical polarization using a Flann 25240-20 Tx horn. The signal then propagated through a Thorlabs PTFE LAT-200 lens, which focused the incident microwave beam in vicinity of the discharge tube axis. The microwave signal scattered by the plasma passed through an identical lens-and-horn arrangement positioned at a 90˚ angle relative to transmitting branch, consisting of a PTFE LAT-200 lens and a Flann 25240-20 Rx horn, and was sent to the RF input of the I/Q mixer. The reference-arm signal passed through a Millitech 45734H-1200 variable attenuator, a Hitachi M4610 variable phase shifter, and a flexible cable before entering the LO input of the I/Q mixer (-3.5 dBm). Output signals of the I and Q channels of the I/Q mixer were recorded using a Lecroy HDO3904 oscilloscope.

A 10 GHz microwave interferometer was used to independently measure the electron number density in the glow-discharge tube. The interferometer operated at 10 GHz and utilized a homodyne detection scheme. The system used an Anritsu 68369B microwave signal generator, a Wiltron D19700 splitter, Mini-Circuits ZX60-183A-S+ amplifiers, and a MITEQ IRM0812LC2Q I/Q mixer. The phase shift introduced by the plasma ($\Delta\phi$) was determined using the mixer, and the average electron number density in the discharge column was calculated as $n_e = \Delta\phi \frac{2c\omega m_e \varepsilon_0}{e^2 L}$, where $c$ is the speed of light, $\omega$ is the angular circular frequency corresponding to 10 GHz, $m_e$ is the mass of electron, $\varepsilon_0$ is the dielectric permittivity of vacuum, $e$ is the electron charge, and $L$ is the plasma path length.

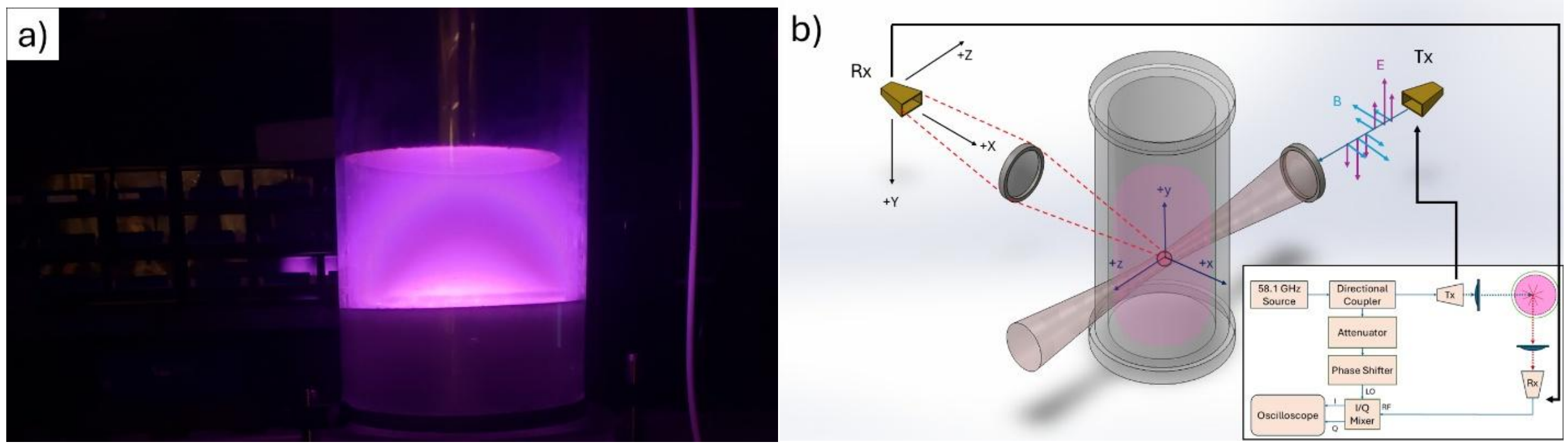

**Figure 1. (a) The DC glow discharge used as the tested plasma object (the positive electrode on top and the grounded electrode on bottom). (b) Schematics of the 58-GHz Focused-CMS system constructed around the glow discharge facility.**

## 3. Results and Discussion

To enable spatial resolution in the CMS technique, two PTFE LAT-200 lenes with a focal length F=20 cm were utilized as follows.

The first lens was utilized to focus the incident microwave beam emitted by the Tx horn and to localize the probing region exposed to significant microwave power to the vicinity of the incident-beam waist, as shown in Figure 1(b). The lens was positioned at a distance of 2F=40 cm from the Tx horn. Correspondingly, the beam waist was formed approximately 40 cm (2F) downstream of the lens. To characterize the focusing performance of the lens, the microwave intensity distribution near the beam waist was measured, and the spatial characteristics of the incident microwave beam, namely the Rayleigh length ($z_R$) and beam waist radius ($w_0$), were evaluated. For this measurement, the circuit shown in Figure 1(b) was modified by bringing the Rx horn near the incident microwave beam waist and scanning it along z-axis ($x = y = 0$). The corresponding signal detected by the Rx horn as a function of the z-coordinate is shown in Figure 2. The plotted quantity is the relative microwave power, $I^2 + Q^2$, where $I$ and $Q$ are outputs of the I/Q mixer. One can see that a clear focal point was observed, corresponding to the maximum measured power. The Rayleigh length was obtained to be $z_R$=14 cm, determined as the axial distance from the focus at which the measured beam intensity decreases to one-half of its maximum value. The corresponding beam-waist radius was then calculated using the Gaussian beam relation: $w_0 = \sqrt{\frac{z_R \lambda}{\pi}}$ =1.52cm, where λ is the probing microwave wavelength (λ =5.16 mm).

The second lens was utilized to map the microwave radiation scattered by the plasma. To this end, the lens was positioned at a distance of 2F (40 cm) from the waist region of the incident microwave beam as shown in Figure 1(b). This configuration produced a 1:1 image of the plasma-scattered microwave field in the plane located at a distance of 2F (40 cm) downstream of the lens. The Rx horn was used to scan the YZ plane shown in Figure 1(b) to reconstruct the image of the scattered microwave field.

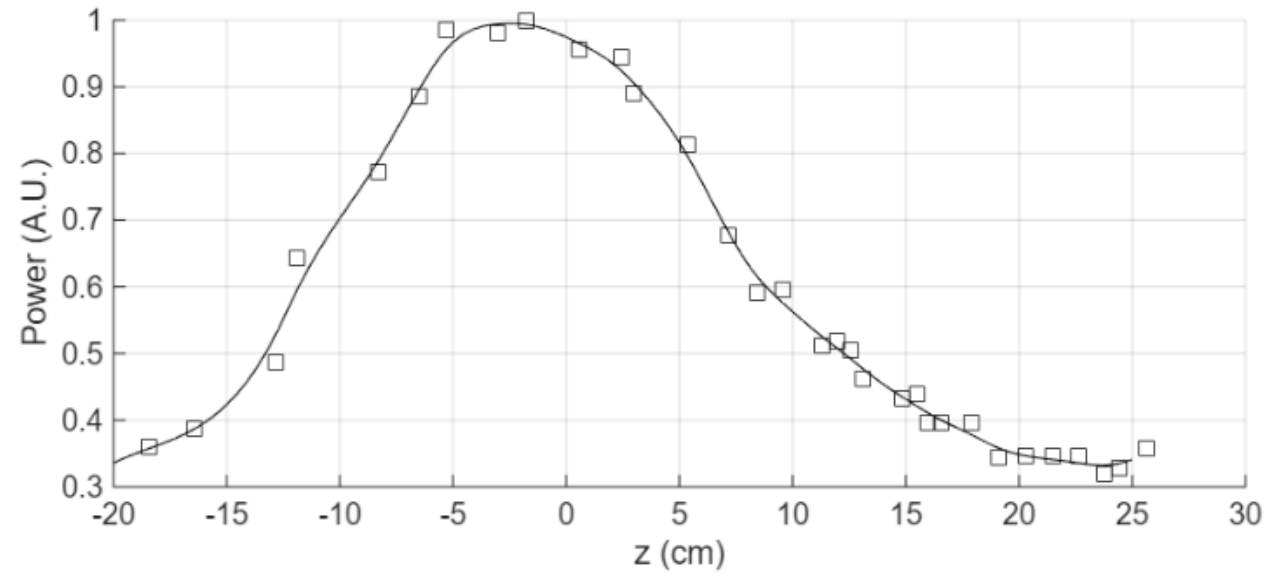

**Figure 2. Incident microwave beam power distribution along z-axis near the beam waist.**

The output signal of the F-CMS system ($V_s$) was determined as follows. With the Tx horn running, and no plasma present in the glow discharge tube, measurements of the I/Q mixer output were taken ($I_0, Q_0$). Then, the discharge was enabled, and measurements were taken again ($I, Q$). Finally, the output signal $V_s$ was calculated as: $V_s = \sqrt{(I - I_0)^2 + (Q - Q_0)^2}$.

The output F-CMS signal ($V_s$) generated by the glow discharge plasma operating at discharge voltage of 630 V and discharge current of 150 mA is shown in Figure 3 (data was collected at 25 points - five Y-coordinates and five Z-coordinates). Note that the average plasma density at these discharge conditions was independently measured using a 10 GHz microwave interferometer and was found to be $n_e$~$5\times10^9$ cm$^{-3}$. One can see that that a clear nonzero scattering signal ($V_s$=13 mV) was observed upon discharge ignition. The spatial characteristics of the scattered microwave field were consistent with those of the incident beam. In particular, the localization of the scattered field in the Y-direction ($\Delta Y_{FWHM}$~30 mm) is consistent with the beam waist diameter $2w_0$~30mm, as expected. The localization of the scattered field in the Z-direction ($\Delta Z_{FWHM}$ ~25 mm) is significantly smaller than $2z_R$~30cm, likely due to the radial variation of $n_e$ in the plasma column and finite size of the discharge tube. Finally, as expected, the magnitude of the $V_s$-signal was observed to increase with increasing discharge current, while the location of the maximum in the YZ plane did not change noticeably.

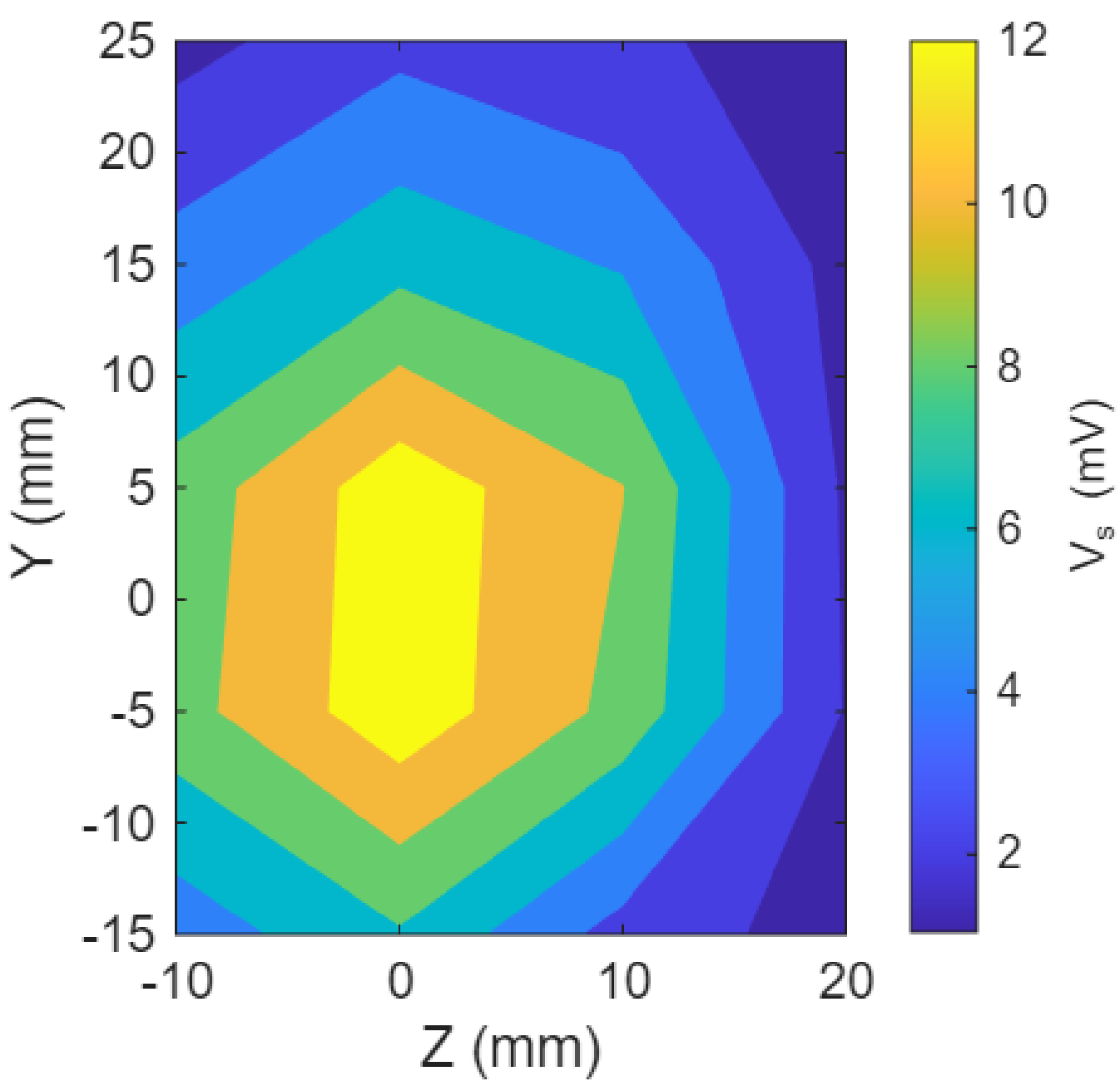

**Figure 3. Contour plot of the output F-CMS signal ($V_s$) generated by the glow discharge plasma column with an average electron number density of $n_e$~$5\times10^9$ cm$^{-3}$.**

Let us now discuss several aspects of the performance and characteristics of the proposed F-CMS diagnostic technique.

One important note is that, in the final implementation of the F-CMS system, mapping of the scattered field conducted in the work and shown in Figure 3 will not be required. Instead, prior to the measurements, the Rx horn will be scanned in the image plane (2F) and then fixed at the location corresponding to the maximum $V_s$ signal during the experiments (this position is associated with the scattered field originating from the center of the incident-beam waist region). The following F-CMS calibration and data acquisition will be conducted with the Rx horn fixed in this position.

The spatial resolution of the F-CMS technique (or equivalently the probing region size) is yet to be verified experimentally; however, it is expected to be on the order of several microwave wavelengths ($\lambda$). The probing-region size in the $x$-direction is governed by the cross section of the incident microwave beam along the line-of-sight and it can be estimated to be on the order of the incident-beam diameter, which was determined in this work to be approximately $2w_0$~30mm. The probing-region sizes in $y$- and $z$-directions were not addressed in this work and will be characterized in future studies. It is important to note that higher microwave operating frequencies (or smaller $\lambda$) can be used if a smaller probing region (finer spatial resolution) is required.

Similar to the conventional CMS technique, the temporal resolution of the F-CMS technique is in the sub-nanosecond range.[2] The temporal resolution of the system utilized in this work was limited by the oscilloscope bandwidth of 3 GHz (approximately 0.3 ns).

The lower detection limit of F-CMS $n_e$-measurements is governed by the sensitivity of the microwave detection system. In this work, an electron number density $n_e$~$5\times10^9$ cm$^{-3}$ was successfully detected; however, it is expected that this sensitivity level can be substantially improved. In general, the sensitivity of the CMS technique is governed by the minimum detectable number of electrons within the probing volume ($N_e$), which is typically on the order of $N_e$~$10^8$ electrons.[2] Therefore, plasmas with very low number densities can be detected by using a larger probing volume, thereby increasing the total number of electrons within it. A larger probing volume can be achieved by reducing the frequency, and therefore increasing the wavelength, of the probing microwaves. Thus, F-CMS allows a tradeoff between spatial resolution and sensitivity; for example, spatial resolution can be sacrificed to improved sensitivity, if needed.

In contrast, the upper detection limit of $n_e$ measurements is linked to the requirement that the scattered signal experiences minimal attenuation as it propagates through the bulk plasma surrounding the probing region toward the Rx horn. The attenuation can be characterized using the factor of $e^{-\alpha L}$, where $L$ is the length of microwave propagation path in the bulk plasma (can be approximated by the plasma column radius of 10 cm in this work) and $\alpha$ is the attenuation coefficient. The attenuation coefficient is defined as $\alpha = -\mathrm{Im}(k)$, where $k = \frac{\omega}{c}\sqrt{\varepsilon}$ is the wavevector in the bulk plasmas. Here, $\omega$ is the angular frequency of the probing microwaves, $c$ is the speed of light, $\varepsilon$ is the dielectric permittivity of plasmas, given by $\varepsilon = 1 - \frac{\omega_p^2}{\omega^2+\nu^2} - i\frac{\omega_p^2}{\omega^2+\nu^2}\frac{\nu}{\omega}$, where $\omega_p$ is the plasma frequency, and $\nu$ is the collisional frequency. For the background air pressure of 0.3 Torr and the plasma column radius of 10 cm considered in this work, the electron number density of $n_e$~$5\times10^9$ cm$^{-3}$ produces attenuation of about 0.002%, which is negligible. Correspondingly, if the maximum acceptable attenuation in the bulk plasma is set to about 10%, the upper measurable limit can be estimated to be $n_e \sim 10^{13}$ cm$^{-3}$.

Finally, the expected performance and key features of the proposed F-CMS diagnostic technique can be summarized as follows. The F-CMS technique enables non-intrusive, absolutely calibrated, spatiotemporally resolved, and highly sensitive measurements of the electron number density (down to $n_e \sim 10^8$ cm$^{-3}$), acquired in real time without the need for signal accumulation and averaging. Similar performance for low-density plasma measurements in real-time can be only provided by the electrostatic Langmuir probes; however, an important advantage of the F-CMS technique is its non-intrusive nature. The F-CMS technique enables sub-nanosecond-scale temporal resolution and is expected to provide adjustable spatial resolution in the millimeter-to-centimeter range, governed by the frequency (wavelength) of the probing microwaves.

These anticipated capabilities of F-CMS enable a wide range of potential applications requiring real-time plasma-density measurements in relatively rarefied plasmas, with $n_e$ in the range of $10^8$-$10^{13}$ cm$^{-3}$. One example is the diagnostics of hypersonic flows, where single-shot measurements are essential because many facilities, such as hypersonic shock tunnels, inherently operate in a single-shot mode.[12] Another important application is diagnostics of plumes of electric propulsion systems, where real-time, temporally resolved measurements are required to accurately characterize unsteady dynamics in relatively rarified plasma plumes.

## 4. Conclusion

This work demonstrates the initial implementation of the Focused Coherent Microwave Scattering (F-CMS) diagnostic technique, a development of the convention CMS technique that offers spatially resolved measurements of electron number density. The feasibility of the approach was confirmed by observing a clear nonzero F-CMS output signal produced by a glow-discharge plasma column with an average electron number density of $n_e \sim 5\times10^9$ cm$^{-3}$. The technique has the potential to enable real-time plasma-density measurements in rarefied plasmas, with $n_e$ down to $10^8$ cm$^{-3}$, a capability that can otherwise be provided only by electrostatic Langmuir probes, while F-CMS offers the important advantage of being non-intrusive. These anticipated capabilities of F-CMS may enable a wide range of potential applications, including measurements in unsteady flowfields such as those occurring in hypersonic shock tunnels and electric propulsion systems.

## Acknowledgments

This work was supported by the National Science Foundation (Grant No. 2409559). We thank Nicholas Babusis and Lee Organski for their assistance with the experimental apparatus used in this work.